\documentclass[reqno]{amsart}

\usepackage{amsmath}
\usepackage{amssymb}
\usepackage{amscd}
\usepackage{mathtools}
\usepackage[utf8]{inputenc}
\usepackage[T1]{fontenc}

\def\dju{\mbox{Đurđevich}}

\theoremstyle{definition}
\newtheorem{definition}{Definition}[section]

\numberwithin{equation}{section}

\begin{document}

	\begin{center} 
		
		{\bf Quantum Probability Geometrically \\
		Realized in 
		Projective Space}
		
		\vskip 0.2cm
		Stephen Bruce Sontz 
		\vskip 0.2cm 
		Centro de Investigaci\'on en Matem\'aticas, A.C.
		\\
		Guanajuato, Mexico
		\\
		email: sontz@cimat.mx 
		\\
		ORCID:  0000-0002-3295-6731
	
	\vskip 0.4cm

    \vskip 0.4cm 
	{\bf Abstract}
		
	\end{center}

\noindent 
The principal goal of this paper is to pass all quantum 
probability formulas 
to the 
projective space associated to the 
complex Hilbert space of a given quantum system, 
providing a more complete geometrization of 
quantum theory. 
Quantum events have 
consecutive and conditional 
probabilities, which have been used 
in the author's 
previous work 
to clarify `collapse' 
and to generalize 
the concept of entanglement by incorporating it 
into quantum probability theory. 
In this way all of standard textbook quantum theory can be
understood as a geometric theory of projective subspaces
without any special role for the zero-dimensional projective 
subspaces, which are also called pure states. 
The upshot is that quantum theory is the probability 
theory of projective subspaces, or equivalently, 
of quantum events.
For the sake of simplicity 
the ideas are developed here in the context
of a type~I factor, but comments will be given
about how to adopt this approach to more general
von Neumann algebras. 


\section{Introduction}

The geometrization of quantum theory is part of a more
general program of the geometrization of physics.
Particular examples are classical mechanics 
viewed as  
symplectic geometry, thermodynamics 
as 
contact geometry, general relativity 
as 
differential geometry and 
gauge theory as 
the geometry of connections on 
principal bundles.  
All of this can be seen as an approach to the sixth
Hilbert problem 
(see \cite{accardi,wightman}), 
which calls for the axiomatization 
of physics. 
Since physics has changed profoundly since 
1900 when Hilbert presented his problems as a challenge 
for the upcoming 20th century, 
quantum physics, unknown in 1900, 
now must be included as part of this axiomatization 
problem. 
It is not a new idea to formulate quantum theory
in a projective space setting, as was already known 
in some sense 
 by
von Neumann in the 1930's. (See \cite{von-neumann}.) 
An extensive treatment was 
given later by Varadarajan in \cite{varad}. 
However, the role of probability in quantum theory 
has not been fully formulated in the 
projective space setting, since the important 
roles of consecutive and conditional probability 
have not yet been geometrized. 
That will be done in this paper, which is a continuation 
of \cite{SBS} where axiomatization is also discussed.

\section{Preliminaries}

In this section we will review known material.  
Some references are \cite{SBS, varad, von-neumann, weidmann}. 
 
Throughout this paper $\mathcal{H}$ denotes a
complex Hilbert
space of any dimension with inner product
$ \langle \cdot, \cdot \rangle$ and norm $|| \cdot ||$. 
We let $\mathcal{L} (\mathcal{H})$ denote the algebra of
all bounded, linear operators 
$T : \mathcal{H} \to \mathcal{H}$. 
We let $ U (\mathcal{H}) $ denote the group of all 
unitary operators in $ \mathcal{L} (\mathcal{H})$. 
For any vector space $V$ we define 
$V^{\mathsf{x}}:= V \setminus \{0\} $. 

The associated {\em projective space} of $\mathcal{H}$ 
is defined as 
$$
\mathbb{C}P (\mathcal{H}):= 
\mathcal{H}^{\mathsf{x}} / \mathbb{C}^{\mathsf{x}},
$$
where one quotients by the action of scalar multiplication of 
$ \mathbb{C}^{\mathsf{x}}  $
on $\mathcal{H}^{\mathsf{x}}$.
Equivalently, 
$\mathbb{C}P (\mathcal{H}) = \mathbb{S}(\mathcal{H})/ \mathbb{S}^{1}$,
where 
$ \mathbb{S}(\mathcal{H}):= \{ \psi \in \mathcal{H} ~|~ ||\psi||=1 \}$
is the {\em unit sphere} in $\mathcal{H}$
and $\mathbb{S}^{1} := \mathbb{S}(\mathbb{C})$ 
is the unit sphere
(or {\em circle}) in the one-dimensional Hilbert space
$\mathbb{C}$. 
This equivalent reformulation works because
Hilbert space has a norm.

It is known that $\mathbb{C}P (\mathcal{H}) $ is a 
K\"ahler manifold, a geometric object with these three compatible
structures: a symplectic form, 
a complex structure and a Riemannian metric. 
This structure as a K\"ahler manifold also holds in the case 
when $\mathcal{H}$ is infinite dimensional, provided one uses 
the appropriate definitions generalizing the usual, finite 
dimensional case. 
(See \cite{ashtekar-schilling, heslot, hughston, kibble, schilling}.)
The points in the geometric space 
$\mathbb{C}P (\mathcal{H})$
are the {\em pure states} 
of the quantum system described by $\mathcal{H}$. 
Note that 
$\mathbb{C}P (\mathcal{H}) = \emptyset$ 
the empty set (or {\em empty space}) if $\mathcal{H} = 0$. 
This trivial case has nonetheless some role to play as we 
shall see. 

We let 
$$
\pi : \mathcal{H}^{\mathsf{x}} \to \mathbb{C}P (\mathcal{H})
$$
denote the quotient map.
It is known that quantum dynamics formulated in 
$\mathcal{H}$ can be ``pushed down'' to a theory
formulated in the geometric space 
$ \mathbb{C}P (\mathcal{H}) $, where it is a
Hamiltonian theory. 
(See \cite{ashtekar-schilling}.)
 However, not all of the probabilistic features  of
quantum theory in the Hilbert space setting 
have been expressed in the purely 
geometric context of 
$\mathbb{C}P (\mathcal{H}) $. 
That is the aim of this paper, which is a continuation 
of \cite{SBS} where the analysis
was done entirely in the setting of $\mathcal{H}$. 

\begin{definition}
For every closed subspace $V$ of $\mathcal{H}$
we say that $\pi (V^{\mathsf{x}})$
is a {\em projective subspace} of $\mathbb{C}P (\mathcal{H})$. 
$\quad \blacksquare$
\end{definition}

Since Hilbert spaces are the primary scenario for quantum theory,
the previous definition is restricted to the case of {\em closed}
subspaces, which are naturally Hilbert spaces in their own right. 

Note that $\pi (V^{\mathsf{x}})$ is clearly
isomorphic to $\mathbb{C}P(V)$ (in the appropriate category). 
For the closed subspace $V=0$, the associated projective 
subspace is empty. 

There is a bijective correspondence between the set 
$ \mathcal{E}$ 
of all
{\em closed} subspaces $V$ of $\mathcal{H}$ and the set $ \mathcal{E}'$ 
of all 
projective subspaces $S$ of $\mathbb{C}P (\mathcal{H})$
given by $V \mapsto \pi (V^{\mathsf{x}})$ 
and $ S \mapsto \pi^{-1} (S) \cup \{0\}$. 
Since $ \mathcal{E}$ is an orthocomplemented, complete
lattice, the same structure passes over to $ \mathcal{E}'$. 
Note that the empty subspace $\emptyset \in \mathcal{E'}$. 
If a projective subspace $S \in \mathcal{E'}$ 
and a closed subspace $V \in \mathcal{E}$ 
correspond to each other in this way, then 
$\dim V = \dim S + 1$, 
provided that we define $ \dim \emptyset := -1$. 

The elements in $ \mathcal{E}$ are also in bijective correspondence
with all the (orthogonal) projection operators $P = P^{2} = P^{*}$
acting 
on $\mathcal{H}$ by the map $ P \mapsto \mathrm{Ran}(P)$, 
the range of $P$. 
Since projection operators are self-adjoint operators 
that represent physical occurrences, they are also called
{\em events} as we will do for the rest of this paper. 
This terminology is explained in the literature. 
(See \cite{beltrametti}, \cite{gudder}, 
\cite{maassen} or \cite{varad} for example.)

Because we now have a bijection between the events
and the projective subspaces, we sometimes call 
the latter {\em geometric events}. 
Of course, {\em physical events}, which are actual occurrences 
in the physical world, are what one has to understand
using a mathematical theory. 

For every  1-dimensional subspace $V$ 
(which is necessarily closed) of 
$\mathcal{H}$, $\pi (V^{\mathsf{x}})$ is a 0-dimensional projective 
subspace (namely, one point) called a pure state. 
In this case for any  $\psi \in V$ which is a {\em unit vector}, 
meaning $|| \psi || =1$, 
we have that $V = \mathbb{C} \psi$ and that the projection operator
whose range is $V$ is $| \psi \rangle \langle \psi |$ in Dirac 
notation\footnote{Definition: 
$ | \psi \rangle \langle \psi | \phi := 
\langle \psi, \phi \rangle \psi$ for all $\phi \in \mathcal{H}$.}. 

In Hilbert space theory, self-adjoint operators are in bijective
correspondence with projection valued measures defined on the
real line $\mathbb{R}$. 
Moreover, the physical observables are represented in quantum theory
by self-adjoint operators.
While there is no natural way in general 
for passing the action of a bounded 
operator on a vector 
to the projective space, something else can be done for
operators having a spectral representation, including 
unbounded self-adjoint operators as is the case in 
the following definition, which already appeared
in \cite{varad}.

\begin{definition}
\label{observable}
A {\em geometric (or quantum) observable} is a map
$$
      E : \mathcal{B} (\mathbb{R}) \to \mathcal{E'}, 
$$
where $\mathcal{B} (\mathbb{R})$ is the $\sigma$-algebra of
Borel subsets of $\mathbb{R}$, satisfying 
\begin{itemize}
	
	\item 
	$ E (\emptyset)  = \emptyset$, \qquad \qquad 
	(The empty set maps to the empty subspace.) 
	
	\item 
	$ E (\mathbb{R}) = \mathbb{C}P (\mathcal{H}) $, 
	
	\item 
	$ E(\cup_{j \in J}  B_{j}) = \sup_{j \in J} E (B_{j}) $ for every 
	pairwise
	disjoint, countable family of Borel subsets 
	$\{ B_{j} \,|\, j \in J \}$
	of $\mathbb{R}$ (with $J$ being a finite or 
	countably infinite index set). $\quad \blacksquare$
	
\end{itemize}

\end{definition}

One could also reasonably call $E$ a {\em projective subspace valued 
measure} on $\mathbb{R}$ or a
{\em geometric event valued measure} on $\mathbb{R}$. 
The third condition justifies the terminology `measure' 
while the first two conditions are normalizations 
that are analogous to those of a probability measure. 

This definition is easily generalized. 
For example, $\mathcal{B} (\mathbb{R})$ could be replaced 
with some other $\sigma$-algebra, or $\mathcal{E'}$ could 
be the subspaces of another geometry. 
Another interesting possibility is to replace $\mathcal{E'}$ 
with the set of only those projective subspaces 
that come from the events in a von Neumann algebra of
operators that act on $\mathcal{H}$. 

We have already realized pure states in the projective space 
setting. 
But one also deals with density matrices 
(also called mixed states) 
in the 
Hilbert space setting.
So we would like to find the corresponding object in  
the projective space 
setting.
A {\em density matrix} in the 
Hilbert space setting is a positive, self-adjoint, 
trace class  
operator with trace $1$. 
This can be transferred over to the projective space setting.
This requires an auxiliary definition as follows. 
\begin{definition}
The {\em support} of a geometric observable 
$E  : \mathcal{B} (\mathbb{R}) \to \mathcal{E'}$
is defined as
$$
\mathrm{supp} (E):= \mathbb{R} \setminus
\bigcup \{ U \subset \mathbb{R}   ~|~ U \mathrm{~is~open~and~}
E(U) = \emptyset \}. \quad \blacksquare
$$
\end{definition}

It immediately follows that $\mathrm{supp} (E)$ is closed, 
but not necessarily bounded. 
If $\mathcal{H} \ne 0$, then it can be shown that
$\mathrm{supp} (E)$ is non-empty.

The spectral theorem for a self-adjoint, positive
trace class operator $T$ says that the spectrum 
of $T $, $\mathrm{Spec}(T)$, is a subset of $[0, \infty)$,  and $\mathrm{Spec}(T) \cap (0, \infty)$ is a bounded,
discrete set with no limit points in
$ (0, \infty)$. 
So $\mathrm{Spec}(T) \cap (0, \infty)$ is a countable sequence, either finite or infinite, of
{\em distinct} positive
numbers $\{ a_{k}\}$ such that each $a_{k}$ is
an eigenvalue of finite multiplicity 
$m_{k} = \dim \mathrm{ker} (a_{k}I - T)$. 
If $T$ has trace $1$ as well, then 
the trace of $T$ is $\sum_{k} m_{k} a_{k} = 1$.  
See \cite{reed-simon,weidmann}.  

This characterization of a density matrix in the 
Hilbert space setting motivates the following 
definition in the projective space setting. 

\begin{definition}
A 
{\em geometric density matrix}, also called a
 {\em geometric mixed state}, is a geometric observable 
$
\rho : \mathcal{B} (\mathbb{R}) \to \mathcal{E'} 
$
satisfying 

\begin{itemize}
	\item 
 $\mathrm{supp} (\rho) \subset [0,\infty)$, 
 and $ \mathrm{supp} (\rho) \cap (0,\infty)$ 
  is a bounded,
discrete set with no limit points in $(0,\infty)$.
\\
Notation: Write $ \mathrm{supp} (\rho) \cap (0,\infty)$ as
a countable sequence, either finite or infinite, of
{\em distinct} positive
numbers $\{ a_{k}\}$.  

\item 
   $\dim \rho(\{ a_{k} \})$ is finite for all $k$. 
   
\item 
    $\sum_{k} 
    \big(1 + \dim \rho(\{ a_{k} \}) \, a_{k}  \big) = 1$. 
$\quad \blacksquare$
\end{itemize}

\end{definition}
There clearly is a bijective correspondence between 
geometric density matrices and 
density matrices in Hilbert spaces.
While geometric 
density matrices can easily be included in the formalism
of this paper, that
entails more cumbersome formulas. 
Therefore, 
they will be only briefly discussed later in this paper. 

The upshot will be that 
standard 
quantum theory 
can be expressed in terms of the projective subspaces arising as the 
values of geometric 
observables and the pure states, which are 0-dimensional 
projective subspaces. 
To achieve that, quantum probability theory must be
fully defined also 
in the geometric setting as we now do. 

\section{Geometric Probability}

In this paper various
probability formulas of quantum theory 
will be established in the setting
of projective geometry. 
The method for doing this is straightforward, since
all the probability formulas in the Hilbert space setting 
(see \cite{SBS})
are functions of events and states, which are in 
bijective correspondence with projective subspaces 
and points, respectively, in the geometric setting. 
Thus, a geometric probability of geometric 
events and geometric states is defined by pulling 
back these objects to the Hilbert space setting 
and then applying the known formula there. 
Or equivalently, one simply notes that the function 
in the Hilbert space setting can be pushed down to
the projective space. 
This definition then becomes an algorithm as well. 

The simplest example is the form of Born's rule for
the probability of a single event $E$, given a 
unit vector $\psi$, in 
the Hilbert space setting. 
(See \cite{SBS}). 
This is: 
\begin{equation}
\label{single-event}
P_{\psi} (E) := || E \psi ||^{2} = 
\langle \psi, E \psi \rangle. 
\end{equation}
Throughout this paper $P_{\psi}$ is used to denote the
various probability functions that depend on a given
unit vector $\psi$.

This procedure might seem to be flawed since the
probability formulas in the Hilbert space setting 
on the right of the defining symbol $:=$
use operations that are not defined in the 
geometric setting. 
In this case those operations are evaluating the action
of an operator on a vector, an inner product 
and a norm\footnote{Strictly speaking, the norm on the unit
sphere in Hilbert space 
passes to projective space, but is the constant function $1$, 
a triviality. 
But $E \psi$ in general is not even in the unit sphere.}. 
In other formulas those operations will be
 composing operators or taking a trace. 
But the results of these formulas will be functions that only
depend on geometric structures. 
It is the overall resulting {\em value} of the function 
that must not depend on these operations.
For 
\eqref{single-event}
the probability function passes down to the projective space 
$\mathbb{C} P (\mathcal{H})$, 
since it only depends on the point $\pi(\psi)$ and 
the projective subspace $\pi ((\mathrm{Ran} \, E)^{\mathsf{x}}  )$ 
in $\mathbb{C} P (\mathcal{H})$, where 
$\mathrm{Ran} \, E$ is the range of $E$. 
So the function itself passes to the projective space.
However, the factorization of the function in the Hilbert
space setting does not pass to a corresponding factorization 
in the projective space. 
It is only the complete composition of the Hilbert space
operations that passes down to the projective space. 
And this will be the case for the rest of the probability
formulas in this paper. 
What is relevant here is the important distinction between
a {\em function} and a {\em formula}. 
The function passes to the projective space, but the 
formula does not. 

Probabilities associated with 
density matrices are handled similarly. 
For example, the probability of a single event $E$, given a 
density matrix $\rho$, in 
the Hilbert space setting is
$
P_{\rho} (E):= 
\mathrm{Tr}  (E \rho)
$,
where $ \mathrm{Tr} $ is the trace of a trace class operator. 
This function passes down to the projective space as well as all of 
the other probability formulas that use density matrices in 
\cite{SBS}. 
But in order to facilitate the flow of ideas these formulas 
will not be discussed further. 

Of course, 
it would be more desirable to have geometric probability  
formulas, namely formulas given directly in terms of those
geometric structures without any reference to 
the overlying Hilbert space. 
On the other hand as already remarked, the Hilbert space formulas provide
useful algorithms for computing these probabilities.

\section{Consecutive Probability}

Consecutive probability is defined as 
the probability of
an {\em ordered} sequence of events, possibly with
repetitions of the same event. 
(See \cite{SBS}.) 
The formula for the consecutive probability 
of finitely many, say $n \ge 1$, events
in the context of Hilbert space is
$$
P_{\psi} (E_{1}, E_{2}, \cdots, E_{n}) = 
|| E_{n} \cdots E_{2} E_{1} \psi ||^{2}, 
$$
where $\psi \in \mathcal{H}$ is
a unit vector and $\{  E_{j} \,|\, j=1,\dots,n \}$ is a time ordered 
sequence of events with $E_{1}$ occurring first, 
$E_{2}$ second and so on. 
We will consider the cases $n=1$ and $n=2$. 
A similar analysis can be made in the general case. 
It seems that the earliest explicit 
reference to this formula as a general rule of
quantum theory is in a paper
\cite{wigner} by Wigner in 1963. 
Consequently, it is often called 
{\em Wigner's Rule.}

We have already seen the case 
of exactly one event $E$ 
(which gives a unique
ordered sequence), 
given a unit vector $\psi$, 
in the previous section. 
To see how this works in more detail, we let
$\{  \alpha_{j}  | j \in J\}$ be {\em any} orthonormal basis of 
$\mathrm{Ran} \, E$,
where $J$ is some index set whose cardinality is equal
to the dimension of $\mathrm{Ran} \, E$. 
(Note that $ J $ is empty when $E=0$, but the following
formulas are valid also in this case since sums indexed by the
empty set are taken to be zero.)
With this notation in place we have that 
$ E= \sum_{j \in J} | \alpha_{j} \rangle \langle \alpha_{j} |$ 
in the strong operator topology (see \cite{reed-simon})
and so we can calculate: 
\begin{equation}
\label{prob-one-event}
	P_{\psi} (E) =  \Big\langle \psi , 
	\sum_{j \in J} | \alpha_{j} \rangle 
	\langle \alpha_{j} , \psi \rangle
	\Big\rangle 
	= 
	\sum_{j \in J} \
	\langle \psi , \alpha_{j} \rangle 
	\langle \alpha_{j} , \psi \rangle
	= 
	\sum_{j \in J} |\langle \alpha_{j} , \psi \rangle |^{2}. 
\end{equation}
By the remarks above, the sum on the right side, 
being equal to the left side, 
is a geometric quantity.
But  
each term $|\langle \alpha_{j} , \psi \rangle |^{2}$ 
in that sum also is a geometric quantity. 
In fact, it is known to be related to the geodesic distance 
between $\pi(\alpha_{j})$ and $\pi(\psi)$ in the
projective space $\mathbb{C} P (\mathcal{H})$. 
(See \cite{schilling}.)
The right side 
is also an analytic quantity that appears
in Bessel's inequality:
$
\sum_{j \in J} |\langle \alpha_{j} , \psi \rangle |^{2}
\le || \psi ||^{2} = 1
$. 

Moreover, we get this graphical representation 
of each term in the sum of the third expression in \eqref{prob-one-event} 
as 
$
\psi \to \alpha_{j} \to \psi
$, 
where $v \to w$ for vectors $v,w \in \mathcal{H}$ represents 
a factor $\langle v , w \rangle$ and a sequence of arrows
indicates multiplication (of complex numbers).
Similarly, the sum in the third expression 
can be represented by
$ 
\psi \to E \to \psi
$. 

Continuing with the case $n=2$, which concerns an ordered pair 
$ E_{1}, E_{2}$ of events and a unit vector 
$ \psi \in \mathcal{H}$, 
the special case of the above 
definition of the consecutive probability 
that $E_{1}$ occurs first
and then $E_{2}$, given $\psi$, is 
\begin{equation}
\label{consecutive-prob}
P_{\psi} (E_{1}, E_{2}) := || E_{2} E_{1} \psi ||^{2}. 
\end{equation} 

Now it might not be completely clear how this formula can 
pass to the projective space 
$\mathbb{C} P (\mathcal{H})$. 
But notice that 
the right side is invariant if we replace 
$\psi$ with $e^{i \theta} \psi$ with $\theta \in \mathbb{R}$. 
So we have a formula that depends only on one state 
and the two projective subspaces associated with
the ranges of $E_{1}$ and $E_{2}$. 
According to the comments in the previous section, 
the function (but not the formula) on the right
side passes to the projective space.  

Let's see what this looks like in detail. 
First, notice that 
the action of a linear, bounded operator
on a unit vector is not necessarily a unit vector
nor even a non-zero vector. 
For example, if $ E_{1} \psi $ is non-zero, then we
can normalize it as
$
E_{1} \psi / || E_{1} \psi ||
$, 
which represents a state that 
is independent of the choice of unit vector $ \psi $
that represents the state in the projective 
space. 
However, typically at least one of these inequalities
is strict: 
$
|| E_{2} E_{1} \psi || \le 
|| E_{1} \psi || \le || \psi ||. 
$
So, there is no obvious 
way to compute $P_{\psi} (E_{1}, E_{2})$
from normalized vectors. 

Also, it is well known that the norm of the Hilbert 
space  passes to the projective 
space trivially. 
It appears that  
the situation is simply hopeless for pushing the 
pieces of this formula down to the projective space. 

Nonetheless, formula \eqref{consecutive-prob}
does pass to the projective space, as we already know. 
To see this in another way 
we use the representations 
$$
   E_{2} = \sum_{j \in J} | \alpha_{j} \rangle \langle \alpha_{j} |  
   \qquad \mathrm{and} \qquad 
   E_{1} = \sum_{k \in K} | \beta_{k} \rangle \langle \beta_{k} |,  
$$
where $\{ \alpha_{j} | j \in J \}$ 
(resp., $\{ \beta_{k} | k \in K \}$) is {\em any} orthonormal basis 
of the range of $ E_{2}$ (resp., $E_{1}$) with 
the sums in the strong operator topology. 
It is important to note these representations of events 
are {\em not} unique for events whose range has dimension
greater than $1$. 
We proceed to calculate 
as follows:
\begin{align}
P_{\psi} (E_{1}, E_{2}) &= 
|| E_{2} E_{1} \psi ||^{2} 
= 
\langle E_{2} E_{1} \psi, E_{2} E_{1} \psi \rangle 
\notag 
\\
&= \sum_{j,k,l,m} 
\Big\langle 
|\alpha_{j} \rangle \langle \alpha_{j} , 
\beta_{k} \rangle \langle \beta_{k} | \psi \,  , \,    
|\alpha_{l} \rangle \langle \alpha_{l} , 
\beta_{m} \rangle \langle \beta_{m} |\psi 
\Big\rangle 
\notag 
\\
&= \sum_{j,k,l,m} 
\langle \beta_{k} , \alpha_{j}  \rangle 
\langle \alpha_{l} , \beta_{m} \rangle
\Big\langle 
|\alpha_{j} \rangle 
\langle \beta_{k} | \psi \,  , \,    
|\alpha_{l} \rangle  
\langle \beta_{m} | \psi 
\Big\rangle 
\notag 
\\
&= \sum_{j,k,l,m} 
\langle \beta_{k} , \alpha_{j}  \rangle 
\langle \alpha_{l} , \beta_{m} \rangle 
\langle \psi, \beta_{k} \rangle 
\langle \beta_{m} , \psi \rangle 
\langle 
\alpha_{j} , \alpha_{l} 
\rangle 
\notag 
\\
&= \sum_{j,k,l,m} 
\langle \beta_{k} , \alpha_{j}  \rangle 
\langle \alpha_{l} , \beta_{m} \rangle 
\langle \psi, \beta_{k} \rangle 
\langle \beta_{m} , \psi \rangle \, 
\delta_{j,l}
\notag 
\\
&= \sum_{j,k,m}  
\langle \psi, \beta_{k} \rangle 
\langle \beta_{k} , \alpha_{j}  \rangle 
\langle \alpha_{j} , \beta_{m} \rangle 
\langle \beta_{m} , \psi \rangle. 
\label{expression}
\end{align}

Now the individual inner products 
in this (possibly infinite, but 
nonetheless convergent) series depend on the choice of
bases. 
However, if 
$\{\tilde{\alpha_{j}} | j \in J \}$ 
(resp., $\{\tilde{\beta_{k}} | k \in K \}$ are other choices for the 
bases, then 
we get the same formula for 
$P_{\psi} (E_{1}, E_{2})$ 
with $\alpha_{j}$ replaced by $\tilde{\alpha_{j}}$ 
and $\beta_{k}$ replaced by $ \tilde{\beta_{k}}$. 
So the form of the expression \eqref{expression} is 
invariant under change of basis. 

A more important property of \eqref{expression} is that
each term in the sum (i.e., each product of 4 inner products)
is invariant under this global phase transformation:
\begin{itemize}
	\item 
	$\alpha_{j} \mapsto \mu_{j} \alpha_{j} $
	
	\item 
	$\beta_{n} \mapsto \nu_{n} \beta_{n} $
	
	\item 
	$\psi \mapsto \lambda \psi$
\end{itemize}
for complex numbers $\mu_{j}, \nu_{n}, \lambda$
satisfying $|\mu_{j}| = |\nu_{n}| = |\lambda| = 1$. 
In other words each term depends only on at most 4 points
(or equivalently, pure states)
$\pi(\alpha_{j})$, $\pi(\beta_{k})$, $ \pi(\beta_{m})$,
$ \pi(\psi) $
in the projective space 
$\mathbb{C} P (\mathcal{H})$, 
while the sum itself depends on all 
(possibly infinitely many)
of these points.
But these are not 4 arbitrary points, 
since either $\pi(\beta_{k}) = \pi(\beta_{n}) $ 
when $k=n$
or
these are two orthogonal\footnote{Even though the inner product
does not pass down to pairs of points in the projective space,
the property of orthogonality does.} 
points, 
$\pi(\beta_{k}) \perp \pi(\beta_{n}) $ when $k \ne n$. 

It is important to emphasize that the factors in each term
of \eqref{expression} do not pass down to the projective 
space, that is, they do not have any geometric significance. 
The same is true for the product  of any proper subset
of the factors in one term. 
However, the entire term (i.e., the product of all the factors)
does pass down to the projective space and 
so
expresses a geometric 
property of the corresponding {\em sequence} of the 
pure states associated to the unit vectors. 
Thus the probability $P_{\psi} (E_{1}, E_{2})$, which we already 
know is a geometric quantity, is expressed in \eqref{expression} 
as a sum of terms, each of which is 
a geometric quantity. 

Nonetheless, the mathematical process for calculating 
each of these terms
(i.e., by evaluating inner products and multiplying them) 
does not correspond to a physical process.
The individual steps in this mathematical 
 process do not have a physical significance; 
they simply form an algorithm whose final outcome does have
physical significance. 
Similarly, the process of summation in \eqref{expression} 
also does not correspond
to any physical event. 
And to suppose otherwise is to take an anthropomorphic view of 
theoretic calculations, namely, that what we do is what 
nature does. 
Of course, what we do may indeed be what nature does, but
that has to be put to experimental test. 
And that entails specifying what the
physical processes are for each computational step and 
how they can be observed. 
As far as I am aware there is no such specification in
the literature for this case. 

Each term in \eqref{expression} can be represented graphically
as
$$
\psi \to \beta_{k} \to \alpha_{j} \to \beta_{m} \to \psi.  
$$
Similarly, the sum of all the terms in \eqref{expression} 
can be represented as
$$
  \psi \to E_{1} \to E_{2} \to E_{1} \to \psi, 
$$
where the choice of the bases is suppressed from the
notation, because the sum is independent of those
choices. 
Actually, we can even take two different orthonormal
bases for the range of $E_{1}$ 
in the second and fourth places 
without altering the sum. 

Summarizing, 
each of the Hilbert space structures $E_{1}, E_{2}$ and $ \psi$ 
in \eqref{expression} 
passes to a geometric structure in the projective space. 
Each event passes to the 
projective subspace determined by its 
necessarily closed 
range
while 
the unit vector passes to a point 
(a 0-dimensional projective subspace). 

Conversely, every projective subspace in 
$\mathbb{C} P (\mathcal{H})$
lifts to a unique
event and 
every point in $\mathbb{C} P (\mathcal{H})$
determines a unique ray or, equivalently,
a unique event with one-dimensional range. 
So, the number 
$P_{\psi} (E_{1}, E_{2})$
is a geometric quantity, which has a probabilistic meaning. 
It is not so clear what a direct geometric 
meaning of this quantity might be, 
though there could be more to say here.

\section{Invariant Functions}

In the previous section we saw how geometric 
properties arise in projective space from probabilistic
properties of quantum theory.
Now those considerations motivate the following definition of 
geometric properties in projective space that may
not come from quantum 
theory. 

Let $n \ge 1$ be an integer and 
let $(p_{1}, p_{2}, \dots , p_{n}) \in 
(\mathbb{C} P (\mathcal{H}))^{n}$, 
the $n$-fold Cartesian product.  
This is a finite, ordered 
{\em sequence} of (not necessarily distinct)
points in the projective space. 
Choose unit vectors $\psi_{1}, \dots, \psi_{n}$ such
that $\pi(\psi_{k}) = p_{k}$. 
Then the {\em amplitude} of this sequence of $n$ points is defined by 
\begin{equation}
\label{define-An}
A_{n}(p_{1}, p_{2}, \dots , p_{n}):= 
    \langle \psi_{1}, \psi_{2} \rangle 
    \langle \psi_{2}, \psi_{3} \rangle 
    \cdots
    \langle \psi_{n-1}, \psi_{n} \rangle 
    \langle \psi_{n}, \psi_{1} \rangle. 
    \end{equation}
This is indeed a function of 
$(p_{1}, \dots , p_{n})$, 
independent of the choices $\psi_{1}, \dots, \psi_{n}$, 
as is 
seen by checking that the right side
does not
change under the transformations 
$\psi_{k} \mapsto \lambda_{k} \psi_{k}$ for $\lambda_{k} \in \mathbb{C}$ satisfying 
$|\lambda_{k}| =1$ for $k= 1, \dots, n$. 
That is, each 
$A_{n}$ is a function reflecting a geometric
property. 
However, a product of a proper subset of the
factors in \eqref{define-An} is not necessarily  
invariant under such a transformation. 
In this regard note the crucial role of the last
factor in \eqref{define-An}. 

More specifically, the projective unitary 
group 
$PU(\mathcal{H}) = U(\mathcal{H})/\mathbb{S}^{1}$ acts 
on the projective space $ \mathbb{C} P (\mathcal{H}) $, 
and so also acts 
coordinate-wise on 
$(\mathbb{C} P (\mathcal{H}))^{n}$.
Then each $A_{n}$
is invariant under this action of $PU(\mathcal{H})$ on 
$(\mathbb{C} P (\mathcal{H}) )^{n}$. 
So the geometry on 
$(\mathbb{C} P (\mathcal{H}) )^{n}$ 
of interest here\footnote{Taking 
the part of the 
Erlangen program that says a geometry of a set is the study of 
all those 
properties invariant under a specific group action on that set.} 
is that 
associated to the group $PU(\mathcal{H})$,
which is smaller than $PGL(\mathcal{H})$ 
and the collineation group, these being 
more common choices for the symmetry group
defining the geometry of $\mathbb{C} P (\mathcal{H})$ and its 
associated spaces. 

For the case $n=1$ we have that $A_{1}(p)=1$ for all
points $p$. 
So in this case we have a trivial invariant. 
However, this is consistent with 
the well known mathematical fact 
that all states are 
unitarily equivalent. 

For $n=2$ we have that 
$$
A_{2}(p_{1}, p_{2}) = |\langle \psi_{1}, \psi_{2} \rangle|^{2} 
\in [0,1], 
$$
which is a well known expression in quantum theory, 
often called the `overlap' of $\psi_{1} $ and $ \psi_{2}$. 
More importantly, 
it also has a probabilistic meaning, since
$$
     P_{\psi_{2}} ( |\psi_{1} \rangle \langle \psi_{1} | ) =  P_{\psi_{1}} ( |\psi_{2} \rangle \langle \psi_{2} | ) =
     |\langle \psi_{1}, \psi_{2} \rangle|^{2}. 
$$

I am not aware of the presence of the invariant 
functions $A_{n}$ in the published mathematical literature 
for $n \ge 3$. 
However, I rather doubt that these could be 
previously unknown.

In general, $A_{n}$ has complex values in the 
closed unit disc, 
$$
\mathbb{D}:= \{   \lambda \in \mathbb{C} ~|~ |\lambda| \le 1 \},
$$
since 
$
| A_{n}(p_{1}, p_{2}, \dots , p_{n}) | \le 1
$
follows from the Cauchy-Schwarz inequality. 
This gives a fibration over $\mathbb{D}$, 
where each 
fiber consists of sequences of points of length $n$
of the same amplitude. 
For example, the fiber over 0 consists of sequences 
for which there is a pair of {\em consecutive} points 
$p_{j}, p_{j+1}$ satisfying 
$p_{j} \perp p_{j+1}$. 
(Arithmetic in the sub-indices is taken 
modulo $n$.)
Notice that each fiber is invariant under the action
by cyclic permutations of the finite cyclic group 
$\mathbb{Z}_{n}$ with $n$ elements. 
So $A_{n}$ passes to the quotient 
$(\mathbb{C} P (\mathcal{H}))^{n}/ \mathbb{Z}_{n} \to \mathbb{D}$. 
In particular, all the expressions in 
\eqref{amplitude-as-expected-value} 
in the next section 
are invariant under
cyclic permutation. 

Under the order reversing map  
$ (p_{1}, \dots , p_{n}) \mapsto (p_{n}, \dots, p_{1})$ 
note that 
$$  
A_{n}(p_{n}, \dots, p_{1}) = 
A_{n}(p_{1}, \dots , p_{n}) ^{*}, 
$$
which says that the order reversing 
map is analogous to time reversal symmetry in 
quantum theory. 

The amplitude $A_{n}$ is not invariant under the action of the
symmetric group $S_{n}$ on $n$ elements for $n \ge 3$.
Here is an example for $n=3$. 
Let $\psi_{1}, \psi_{3} \in \mathcal{H}$ be linearly independent 
unit vectors 
with
$\lambda := \langle \psi_{1}, \psi_{3} \rangle \ne 0$.  
(This supposes that $\dim \mathcal{H} \ge 2$.)
Define $\psi_{2}:= (\psi_{1} + \psi_{3}) / N $, where
$N:= || \psi_{1} + \psi_{3} ||$. 
Consider the points $p_{j}:= \pi(\psi_{j})$ for $j=1,2,3$. 
Then we have
$$
A_{3} (p_{1}, p_{2}, p_{3}) = \langle \psi_{1}, \psi_{2} \rangle 
\langle \psi_{2}, \psi_{3} \rangle
\langle \psi_{3}, \psi_{1} \rangle = \dfrac{1}{N} ( 1 + \lambda)
\dfrac{1}{N} ( 1 + \lambda) \lambda^{*}, 
$$
but on the other hand
$$
A_{3} (p_{1}, p_{3}, p_{2}) = \langle \psi_{1}, \psi_{3} \rangle 
\langle \psi_{3}, \psi_{2} \rangle
\langle \psi_{2}, \psi_{1} \rangle = 
\lambda \,  \dfrac{1}{N} ( 1 + \lambda^{*}) 
\dfrac{1}{N} ( 1 + \lambda^{*}). 
$$
These two expressions are not equal if furthermore we choose
$\psi_{1}, \psi_{3}$ such that $\lambda \notin \mathbb{R}$ and
$|\lambda| < 1$. 

For the case $n > 3$ we consider
$$
A_{n} (p_{1}, p_{2}, p_{3}, p_{1}, p_{1}, \dots, p_{1})
\qquad \mathrm{and} \qquad  
A_{n} (p_{1}, p_{3}, p_{2}, p_{1}, p_{1}, \dots, p_{1}),
$$
each of which reduces to the formula for the case $n = 3$. 
It seems fair to say that 
this geometric fact is related to a quantum effect, since it shows 
that order is important. 

This same example shows that
for $n\ge 3$ the function $A_{n}$ can assume non-real 
values and so this value does not have a physical 
significance, though it does have a geometric significance. 
However, $|A_{n}|^{2}$ takes values in $[0,1]$ and therefore
could have a physical significance.\footnote{The exponent $2$
is in resonance with the relation in quantum theory between a
probability amplitude and its associated probability. 
But other values could
be used.}

A curious special case of this construction arises in
quantum probability, namely for a palindrome of an odd number
$2 k - 1$  of points: 
$$
p_{1}, \dots , p_{k-1}, p_{k} , p_{k-1}, \dots , p_{1}. 
$$
In this case the amplitude of this sequence lies in $[0,1]$. 

Everything in this section works for (possibly incomplete)
inner product spaces over $\mathbb{C}$, 
and it all holds for other scalar fields, most notably $\mathbb{R}$, 
as well.

\section{Integrals, Expected Values, States, Symbols, Forms}

We now note that $A_{n}$ has a mathematical significance, 
which has been given a wide variety of names. 
(We assume $\mathcal{H} \ne 0$ in this section.)
First we define something that comes from
mathematical analysis, namely the {\em integral}.  
The integral $\mathbb{E}_{p}$
is defined for a state $p \in \mathbb{C} P (\mathcal{H})$ 
represented by a unit vector
$\psi \in \mathcal{H}$ and a bounded,
but not necessarily self-adjoint,
 operator 
$T \in \mathcal{L}(\mathcal{H})$ as
$$
\mathbb{E}_{p} (T):= \langle \psi, T \psi \rangle 
\in \mathbb{C}, 
$$
which is well defined since the right side only depends
on $p = \pi(\psi)$. 
This notation emphasizes that
$\mathbb{E}_{p} : \mathcal{L}(\mathcal{H}) \to \mathbb{C}$
 for a given $p$ is a 
linear functional $ \mathbb{E}_{p} $, which is also called a {\em state} 
in the theory of $C^{*}$-algebras. 
This integral $ \mathbb{E}_{p} $   also has
a probabilistic significance, which is reflected in its being dubbed the
{\em expected value (in the state~$p$)} as well. 
This nomenclature explains the origin of its notation. 

But we can also take a fixed $T \in \mathcal{L}(\mathcal{H}) $ 
and define a
function $\sigma(T): \mathbb{C} P (\mathcal{H}) \to \mathbb{C}$ 
by reversing roles:  
$$
\sigma(T) (p) := \mathbb{E}_{p} (T) = \langle \psi, T \psi  \rangle \qquad \mathrm{for~} p = \pi(\psi) \in 
\mathbb{C} P (\mathcal{H}) \mathrm{~with~} || \psi || = 1. 
$$
This mapping $T \mapsto \sigma(T)$ is a type of 
dequantization,\footnote{``Functions instead of operators''
whereas quantization is ``operators instead of functions''.} 
but it is also a type of symbol of the operator. 
It is called a {\em covariant symbol} or a {\em lower symbol}. 
The latter name comes from the property that $\sigma(T)$ is a
bounded function which satisfies
$$
     || \sigma(T) ||_{\infty} := 
     \sup_{p \in \mathbb{C} P (\mathcal{H})}
     | \sigma(T) (p) | = 
     \sup_{||\psi||=1} |\langle \psi, T \psi \rangle| 
     \le \sup_{||\psi||=1} ||\psi||^{2} || T || =
     || T ||,
$$
that is, the lower symbol of $T$ provides a lower bound 
on the operator norm of $T$. 

Now $A_{n}$ can be expressed in terms of these constructions as follows: 
\begin{equation}
	A_{n}(p_{1},  \dots , p_{n}) = 
	\mathbb{E}_{p_{1}} (|\psi_{2}\rangle \langle \psi_{2}|
	\cdots |\psi_{n}\rangle \langle \psi_{n}| ) =  
	\sigma \big(|\psi_{2}\rangle \langle \psi_{2}|
	\cdots |\psi_{n}\rangle \langle \psi_{n}|\big) (p_{1}). 
	\label{amplitude-as-expected-value}
\end{equation}

Defining 
$ \mathcal{F} := \{ f: \mathbb{C} P (\mathcal{H}) \to \mathbb{C} ~|~ f \mathrm{~is~bounded} \} $, we have that
$\sigma : T \mapsto \sigma(T)$ is a mapping
$\sigma : \mathcal{L}(\mathcal{H}) \to \mathcal{F}$
which is linear and satisfies 
$\sigma (T^{*}) = \sigma(T)^{*}$. 
Consequently, $ \sigma(T) $ is real valued if $T$ is
self-adjoint, that is, if 
$ T=T^{*}$. 

It turns out that an application of the polarization identity shows 
that $\sigma$ is injective. 
In some sense this means that knowing what $\sigma(T)$ is 
tells us what $T$ is. 
This point of view is advocated in \cite{ashtekar-schilling} and
\cite{schilling} in order to construct a Hamiltonian formalism 
of quantum theory based on the symplectic manifold 
$\mathbb{C} P (\mathcal{H})$, which plays the role of a 
phase space. 
In that formalism the symbol $\sigma(T)$ of a 
self-adjoint, bounded operator is 
a real valued, bounded function 
$ \sigma(T) : \mathbb{C} P (\mathcal{H}) \to \mathbb{R}$
defined on the phase space. 
However, not all real valued, bounded functions 
$ f : \mathbb{C} P (\mathcal{H}) \to \mathbb{R}$ arise
as symbols. 
So, self-adjoint, bounded operators are encoded geometrically 
as certain, but not all, real valued, bounded functions. 
The other way given 
in Definition~\ref{observable} 
of representing self-adjoint operators 
as geometric observables has the property that this is a bijective 
correspondence. 
I am not arguing that the approach in this paper 
is better, but rather 
that there are two quite different ways for passing 
self-adjoint operators to the geometric setting. 

As if there were not enough distinct ways of considering 
the expression $ \langle \psi, T \psi  \rangle $, it must 
be noted that it is also called a 
{\em diagonal matrix element of $T$}  
in physics. 
This is a special case of a {\em matrix element} defined
as $ \langle \phi, T \psi  \rangle $ for {\em any} pair 
$\phi, \psi \in \mathcal{H}$. 
Now the expression 
$B(\phi,\psi):= \langle \phi, T \psi  \rangle $ defines a {\em bounded,
sesquilinear form}. 
Such forms are in bijective correspondence with bounded linear
operators acting in $\mathcal{H}$. 
(See for example \cite{reed-simon} for definitions and proof.
Also, $ \langle \psi, T \psi  \rangle $ is called a {\em bounded, 
quadratic form}, yet another name for this expression.)
The upshot is that it is possible to reduce the study of bounded
linear operators to that of bounded, sesquilinear forms. 
But historically the reverse happened. 
Originally D.~Hilbert and his collaborators studied forms. 
The advantage of using operators instead of forms
becomes apparent when trying to understand other
properties (such as composition of operators 
or the resolvent of
an operator) 
in terms of forms. 
This change to the modern viewpoint of thinking in terms of
operators, and eventually operator algebras, comes 
from work of I.~Fredholm and F.~Riesz. 
(See \cite{blackadar}, p. 7.)
The upshot of this is that even though every bounded linear
operator can be passed to a function on the projective space,
namely its lower symbol, getting results is more readily done in
the Hilbert space setting where more mathematical tools are
available. 
In this regard the theory of von Neumann algebras 
provides a large collection
of such tools.

\section{Quantum Theory without States}

The projective space 
$\mathbb{C} P (\mathcal{H})$ 
provides a geometric setting for
understanding quantum theory.  
As one example, a 
consecutive probability is a 
function taking an ordered sequence of 
projective subspaces and one point 
into a number in $[0,1]$. 
That point is a pure state, and it appears to 
have a special role. 
However, that point is also a 0-dimensional 
projective 
subspace and can be viewed as the event
$ E_{0}:= |\psi \rangle \langle \psi|$, 
where $\psi \in \mathrm{Ran} \, E_{0}$ satisfies $|| \psi || = 1$. 
(Notice that the existence of such an event implies that
$\mathcal{H} \ne 0$.)
Here is a computation using this idea, where $|| \cdot ||_{op}$ 
is the operator norm: 
\begin{align*}
|| \, E_{n} \cdots E_{2} E_{1} E_{0} \,||_{op}
&= || \, E_{n} \cdots E_{2} E_{1} |\psi \rangle \langle \psi|\,||_{op}
\\
&= 
\sup_{||\phi||=1} 
||\, E_{n} \cdots E_{2} E_{1} |\psi \rangle \langle \psi|\phi \,||
\\
&= 
\sup_{||\phi||=1} 
||\, E_{n} \cdots E_{2} E_{1} \langle \psi,\phi \rangle \psi \,||
\\
&=
\sup_{||\phi||=1} 
| \langle \psi,\phi \rangle | \, 
||\, E_{n} \cdots E_{2} E_{1} \psi \,||
\\
&=
||\, E_{n} \cdots E_{2} E_{1} \psi \,||
\end{align*}
This shows that a consecutive probability is
given in terms of the operator norm of an 
ordered product of events, namely
$$
 P_{\psi} (E_{1}, E_{2}, \cdots, E_{n}) = 
 ||\, E_{n} \cdots E_{2} E_{1} \psi \,||^{2} = 
 || \, E_{n} \cdots E_{2} E_{1} E_{0} \,||_{op}^{2}, 
$$
where 
$ E_{0}:= |\psi \rangle \langle \psi|$. 
The next definition generalizes this formula. 
\begin{definition}
Given an ordered sequence of $n+1$ 
events, say $E_{0}, E_{1}, \cdots, E_{n}$, we
define the {\em consecutive probability} 
of their occurring in that order as
$$
P (E_{0}, E_{1}, \cdots, E_{n}) := 
|| \, E_{n} \cdots E_{2} E_{1} E_{0} \,||_{op}^{2}. 
\quad \blacksquare
$$
\end{definition}
\noindent 
{\bf Remarks:}
The simplest case in standard quantum theory is
the probability of one measurement, that is one
event $E_{1}$, occurring after the state associated
to a unit vector $\psi$ has been prepared. 
That probability is equal to $P (E_{0}, E_{1})$, 
where $E_{0}$ is given above. 
The probability of one single event (for example, one measurement)
occurring without any given prior or subsequent event or measurement 
is not ever a consideration in scientific practice or discourse. 
So the fact that $ P(E)= 1$ for any event $E \ne 0$ is not
relevant to any experimental or observational event, which is
never considered as isolated from all other events. 
Here the relevant comment is: 
{\em science is based on correlations}. 

However, the formula in the definition above makes perfect sense for
an event $E_{0}$ that has range $\ge 2$, in which case it is 
not represented by a unit vector. 
In such a case we do not have a situation where 
a ``state has been prepared'', as is usually said. 
Nonetheless, this more general situation in which
an ``event has been 
prepared''\footnote{Of course, we could also 
simply say ``an event has occurred''.}
does aptly correspond 
to certain physical conditions. 
One such example is a system which is initially 
constrained spatially to, say, a non-empty open
set. 
In that case 
the initial event $E_{0}$ would be a projection 
operator with infinite dimensional range. 

Now the time evolution 
for a sequence of events $E_{0}, E_{1}, \dots , E_{n}$ occurring at 
times $ t_{0} < t_{1} < \cdots  < t_{n-1} < t_{n} $
is given by the time dependent
Wigner's Rule (see \cite{wigner}), also called
the generalized Born's Rule in \cite{SBS}, 
\begin{align*}
P (E_{0}, E_{1}, \cdots, E_{n-1} , E_{n} &;
t_{0}, t_{1}, \cdots , t_{n-1}, t_{n}) := 
\\
&|| \, E_{n} U_{n,n-1} E_{n-1} \cdots E_{2} U_{2,1} E_{1} U_{1,0} E_{0} \,||_{op}^{2}, 
\end{align*}
where one can take unitary operators 
$U_{k,k-1} = e^{-i (t_{k} - t_{k-1} ) H}$
for some self-adjoint operator $H$, though there are
other possibilities. 
Actually, the fact that 
unitary operators on the Hilbert space pass down to the 
projective space is irrelevant here, since other 
parts\footnote{These being composition of operators and
the operator norm.}
 in this formula do not pass down. 
For this reason one need not take $ U_{k,k-1} $
to be a unitary operator. 
However, to get a probability 
one should restrict these operators to be
contractions, namely, $|| U_{k,k-1} || \le 1$, so that
the formula above gives a number in $[0,1]$. 
On the other hand, the left side is a geometric 
quantity even for more general choices of the 
operators $ U_{k,k-1} $. 

A useful fact is that for any bounded operator $A$ and
any unitary operator $U$ we have that
$$
    || U A ||_{op} = || A ||_{op} \qquad \mathrm{and}
     \qquad || A U ||_{op} = || A ||_{op}. 
$$
This implies that putting a unitary operator to the
right of $E_{0}$ (corresponding to some time evolution
prior to $E_{0}$) has no effect on the overall probability.
This can be thought of as a Markovian type property, namely 
what happens subsequent to the event $E_{0}$
depends only on the event $E_{0}$ and not 
the details of the recent 
prior time evolution. 

Similarly, a unitary operator to the left of $E_{n}$  
has no impact on the probability. 
However, the time evolutions in the time periods between 
events do matter. 

As discussed later in Section~9, 
all quantum probabilities can be defined in terms of
consecutive probability. 
Thus the upshot of this section is that 
standard text book quantum
theory can be formulated in terms of events 
(equivalently, projective subspaces) 
without any special role being played by the pure states
(equivalently, zero-dimensional projective subspaces). 
More simply put, all of quantum theory, 
including all probabilities, 
is a geometric theory of a
projective space and those of its projective 
subspaces coming from the events in a von Neumann algebra.

\section{Quantum Interference}

It is quite remarkable that quantum interference effects
pass to the projective space setting. 
In other words quantum interference is a 
geometric property. 
Let $E_{j}$ be a countable, disjoint family of events,
$E$ an event and 
and $\psi$ a unit vector. 
In this case 
$\sum_{j} E_{j} = \sup_{j} E_{j} $, 
where the (possibly infinite) sum converges in 
the strong operator topology to 
the {\em event}\footnote{The point here is that 
$\sum_{j} E_{j}$ in general is not an event.} 
$\sup_{j} E_{j}$. 
(See \cite{kadison}.)
Then we have: 
\begin{align*}
	&P_{\psi} ( \sum_{j} E_{j} , E ) = 
	P_{\psi} ( \sup_{j} E_{j} , E ) 
	= || E (\sup_{j} E_{j}) \psi||^{2} 
	= 
	\langle 
	E (\sum_{j} \! E_{j}) \psi,
	E (\sum_{k}\! E_{k}) \psi 
	\rangle 
	\nonumber
	\\
	&=
	\sum_{j,k}  
	\langle 
	E E_{j} \psi,
	E E_{k} \psi 
	\rangle 
	= \sum_{j}
	|| E E_{j} \psi ||^{2} + 
	\sum_{j \ne k} 
	\langle 
	E E_{j} \psi,
	E E_{k} \psi 
	\rangle 
	\\
	&= \sum_{j}
	P_{\psi} ( E_{j} , E ) 
	+ 
	\sum_{j \ne k} 
	\langle 
	E E_{j} \psi,
	E E_{k} \psi 
	\rangle.  
\end{align*}

The expression on the far left is a function
of structures that pass down to the projective space.
More noteworthy is that the terms in the sums on the 
far right are also functions that pass down to 
the projective space. 
This is already known for the first sum, since all those
terms are probabilities.
But each {\em interference term} (namely, those in the $j \ne  k$
sum) also is a geometric quantity. 
This is more remarkable, since these interference terms
can be complex numbers that are not real. 
So such terms are not probabilities. 

This identity shows how $\sigma$-additivity is
violated in quantum probability by the presence of the 
interference terms. 
Since this result is a valid identity in 
quantum probability, it is called
{\em quantum $\sigma$-additivity}. 
See \cite{SBS} for more details.

\section{Collapse, Entanglement and All That}

The essential, unifying concept for this section is
{\em conditional probability}, which is defined 
in terms of consecutive probabilities. 
(See \cite{beltrametti} and \cite{SBS}.)
The simplest, but completely illustrative, case is 
given in the following definition. 
\begin{definition}
Given an ordered pair of events $E_{1}, E_{2}$ and a unit vector
$\psi$, the {\em conditional probability} of $E_{2}$ given 
the prior occurrence of $E_{1}$ is defined as
\begin{align*}
P_{\psi} (E_{2} | E_{1}) &:=
\dfrac{P_{\psi} (E_{1}, E_{2})  }{P_{\psi} (E_{1})} 
= \dfrac{|| E_{2} E_{1} \psi ||^{2}}{|| E_{1} \psi ||^{2}}
\qquad 
\mathrm{provided~that~} P_{\psi} (E_{1}) \ne 0,
\\
 P_{\psi} (E_{2}  |  E_{1})&:= 0 \qquad \mathrm{otherwise}. 
\quad \blacksquare
\end{align*}
\end{definition}

Recall that $P_{\psi} (E_{1})$ is a function that passes to
the projective space.  
So the dichotomy in this definition depends on a 
projective space property. 
Also, each of the two possible formulas passes to the
projective space. 
In fact, the first alternative is a formula whose 
numerator and denominator individually pass to the
projective space, while the second formula being a constant 
trivially passes. 
This might seem strange, since the vector $E_{1} \psi \in \mathcal{H}$ does not pass to anything in  the
projective space in general. 
This is simply another manifestation that a function is
not the same thing as a formula. 

Since conditional probability passes to the projective
space, 
it thereby acquires a geometric significance. 
Since `collapse' and entanglement are defined in terms of
conditional probability in \cite{SBS}, these then also become 
understandable in terms of the
geometric projective space. 

For the case of entanglement one first tries defining independence 
in analogy with classical probability theory 
as $ P_{\psi} (E_{2}  |  E_{1}) = P_{\psi} (E_{2} )$
for $\psi$ and $E_{1}, E_{2}$ as in the definition above. 
Since this attempt does not handle well 
the case $P_{\psi} ( E_{1} )=0$, 
the condition (obtained by `cross multiplying')
$$
P_{\psi} (E_{1}, E_{2}) = P_{\psi} (E_{1}) P_{\psi} ( E_{2}) 
$$
is defined to be {\em quantum independence} of the ordered 
pair $E_{1}, E_{2}$, given $\psi$. 
Clearly, this condition passes to the projective
space. 
Since the left side of this condition is not invariant 
in general under the interchange of $E_{1} $ and $ E_{2}$, 
while the right side is invariant, it follows that 
a pair of events can be quantum independent in one order but not in
the other order. 
Finally, {\em entanglement} of an ordered 
pair $E_{1}, E_{2}$, given $\psi$, is defined as the
negation of this property, namely as 
not quantum independence. 
Consequently, entanglement is a geometric property. 

A common way 
to calculate the conditional probability is to use these
identities if the condition 
$  P_{\psi} (E_{1}) = || E_{1} \psi ||^{2} \ne 0 $ is
satisfied\footnote{This translates into English as:
The event $E_{1}$ can occur given the state determined by $\psi$.}:
$$
P_{\psi} (E_{2} | E_{1}) 
= \dfrac{|| E_{2} E_{1} \psi ||^{2}}{|| E_{1} \psi ||^{2}}
= \Big| \Big| \dfrac{1}{|| E_{1} \psi||} E_{2} E_{1} \psi 
\Big| \Big|^{2}
= || E_{2} \tilde{\psi} ||^{2},
$$
where $ \tilde{\psi} := E_{1} \psi/ || E_{1} \psi||$ is a 
unit vector. 
The mapping $\psi \mapsto \tilde{\psi}$ is commonly called
the `collapse of the wave function'.
It is seen to be one step in an algorithm to compute 
$|| E_{2} \tilde{\psi} ||^{2}$, which is equal to 
$ P_{\psi} (E_{2} | E_{1}) $. 
Not all of 
the individual steps in this algorithm pass to the projective
space. 
For example,  
the action of $E_{2}$ on $ \tilde{\psi} $
and the operation of taking the norm do not pass. 
Nonetheless, the function does pass to the projective
space. 
This analysis does not invalidate the `collapse' algorithm, 
but highlights what that algorithm is computing in terms of
the geometry of projective space and quantum probability. 
As emphasized earlier, the individual steps in an 
algorithm for computing a physical quantity 
do not necessarily have any physical significance.  
It is the result of that algorithm which must 
confront experiment. 

\section{Conclusion}

The exact 
geometric significance 
of all these invariants arising from quantum 
probability, including
all the probability functions and especially certain terms in
those formulas, has quite 
possibly not been fully elaborated, most particularly in
infinite dimensions. 
The invariant functions $A_{n}$ may be known 
in invariant theory of classical groups, though perhaps
not in infinite dimensions. 

Moreover, the idea
behind $A_{n}$ can be extended to infinite sequences
of points, indexed by all the integers
(including negative integers), provided that the 
resulting infinite product converges. 
This latter construction 
is possibly something new and merits further 
consideration. 
Another generalization is to define
\begin{equation*}
	A_{n,\sigma}(p_{1}, p_{2}, \dots , p_{n}):= 
	\langle \psi_{1}, \psi_{\sigma(1)} \rangle 
	\langle \psi_{2}, \psi_{\sigma(2)} \rangle 
	\cdots
	\langle \psi_{n}, \psi_{\sigma(n)} \rangle,
\end{equation*}
where $\sigma$ is any permutation of
$\{ 1, \dots , n \}$. 
In this paper we only considered the case when
$\sigma$ is an $n$-cycle. 

While everything in this paper was done in the projective
space associated to a Type~I factor, this method should be 
applicable to other von Neumann algebras. 
Choosing a von Neumann algebra is 
a question of finding the appropriate quantum model 
for a given quantum system. 
The idea is to consider 
the geometry of only those 
projective subspaces 
of $\mathbb{C} P (\mathcal{H})$ which come from the 
events in a von Neumann algebra of operators acting in 
$\mathcal{H}$. 
Those events ($\equiv$ projections) in the von Neumann algebra
should exactly encode the physical occurrences in the quantum system. 
As such, this is a problem of finding the correct 
{\em quantization}. 
This would then provide the geometric setting corresponding to 
an analytical structure.
However, it is not to be expected that the geometrical 
setting will usurp the role of the analytical 
von Neumann algebra or of analysis in general in the Hilbert space setting. 
Far from it. 
Actually, it is quite possible that geometrical results  
could feed back 
to give new insights 
to the analytical setting.

The points in projective space are also in bijective correspondence
with rank~1 projections, which in turn all lie in the
Hilbert space of Hilbert-Schmidt operators. 
This provides an analytic approach,
which has not been used here, for studying the 
projective space $\mathbb{C} P (\mathcal{H})$.

\vskip 0.4cm
\begin{center}
	ACKNOWLEDGMENTS
\end{center}

\noindent
I thank Gil Bor for several enlightening comments
and  Micho \dju~
for years of helpful conversations.

\end{document}